\documentclass{icrc2009}

\usepackage{graphicx}   
\usepackage[caption=false]{caption}    
\usepackage[font=footnotesize]{subfig} 
\usepackage{fixltx2e}
\usepackage{textcomp}
\usepackage{url}

\newcommand{\shorttitle}[1]%
{\markboth{Proceedings of the 31\MakeLowercase{$^{st}$} ICRC, {\L}\'{o}d\'{z} 2009}{#1} }
\newcommand{\etal}{\MakeLowercase{\textit{et al. }}} 

\begin{document}
\title{Multiwavelength observation of the blazar 1ES1426+428 in May-June 2008}

\author{\IEEEauthorblockN{E. Leonardo,\IEEEauthorrefmark{1}
                          D. Bose,\IEEEauthorrefmark{2}
                          N. Mankuzhiyil,\IEEEauthorrefmark{3}
                          I. Oya,\IEEEauthorrefmark{2}
                          S. R\"ugamer,\IEEEauthorrefmark{4}\\
                          A. Stamerra,\IEEEauthorrefmark{1}
                          F. Tavecchio \IEEEauthorrefmark{5}\\
                        on behalf of the MAGIC collaboration}
 L. Foschini \IEEEauthorrefmark{5}and G. Tagliaferri \IEEEauthorrefmark{5}
\IEEEauthorblockA{\IEEEauthorrefmark{1}Universit\`a di Siena and INFN Pisa, I-53100 Siena, Italy}
\IEEEauthorblockA{\IEEEauthorrefmark{2}Universidad Complutense, E-28040 Madrid, Spain}
\IEEEauthorblockA{\IEEEauthorrefmark{3}Universit\`a di Udine and INFN Trieste, I-33100 Udine, Italy}
\IEEEauthorblockA{\IEEEauthorrefmark{4}Universit\"at W\"urzburg, D-97074 W\"urzburg, Germany}
\IEEEauthorblockA{\IEEEauthorrefmark{5}INAF National Institute for Astrophysics, I-20121 Milano, Italy}
}

\shorttitle{E.Leonardo \etal MWL observation of 1ES1426+428.}
\maketitle

\begin{abstract}
We performed a simultaneous multiwavelength observation of the blazar 1ES1426+428 in May-June 2008 with the KVA telescope for the optical band, the Suzaku and Swift satellites for the X-rays and with the MAGIC telescope for the very high energy (VHE) band (E $>$ 100~GeV). \\
 The analysis of the X-ray data showed that the source was in an intermediate state of activity, while MAGIC could not detect significant VHE gamma-ray emission from the source. This indicates that the VHE flux was more than one order of
magnitude lower compared to historical detections. The optical flux increased by ~20 \% during the MAGIC observations.\\
 We present an upper limit at 95 \% confidence level (CL) of 3 \% of the Crab Nebula for E $>$ 130~GeV from about 34 hours of MAGIC observations and the results from Suzaku, Swift and KVA. \\Moreover an SED is obtained combining the multiwavelength observations of the source.\\
  \end{abstract}

 
\section{Introduction}
  Blazars are radio-loud active galactic nuclei whose emission originates from a relativistic jet closely aligned with our line of sight. BL Lac objects, a subclass of blazars, are completely dominated by nonthermal emission and have a strong continuum stretching from radio frequency through $\gamma$-ray frequencies. They also show prominent variability at all wavelengths, with flux changes by several order of magnitude over a few days or weeks.

The multiwavelength spectral energy distribution of high frequency-peaked BL Lac objects (HBL) shows the presence of two well defined broad components, the first one peaking in the UV/X-ray band and the second one in the GeV-TeV band. The lower energy peak is attributed to synchrotron emission by relativistic electrons within the jet, while, according to Synchrotron-Self Compton models \cite{GM}, the second component is due to Inverse Compton emission (IC) from the same electron population scattering the synchrotron photons. In External Compton models ambient photon fields provide additional targets for the IC process \cite{EC}.

Using VHE observations to determine the position of the second peak of the SED allows us to determine associated parameters of emission models and thus to constrain the maximum energy to which particles responsible for the emission are accelerated.

Amongst BL Lac objects, extreme blazars are very interesting objects to be observed by ground-based $\gamma$- ray telescopes. The term "extreme blazars" was introduced in \cite{Ghis1999} to describe those BL Lac objects whose first peak sometimes has been located occasionally in the hard X-ray band. Such objects, which lie at the end of the "blazar sequence" proposed in \cite{Fossati97}, are good candidates for TeV emission since the second peak of their SEDs is supposed to lie at higher energies, meaning that they would emit more power in the VHE regime.

1ES1426+428 \cite{Whipple} is an extreme HBL with the synchrotron peak occasionally observed at E $>$ 100 keV \cite{Wolt}. Its relatively high redshift (z=0.129) makes particularly interesting the study of the absorption of the TeV flux due to the $ \gamma \gamma \to e^+ e^-$ process with the photons of the Infrared Extragalactic Background Light (IR EBL).

\section{MWL OBSERVATION} 
  \subsection {VHE band observations with MAGIC} 
 MAGIC is a single dish Imaging Atmospheric Cherenkov Telescope with a 17 m-diameter main reflector located in the Canary Island of La Palma, operating since 2004 with a low energy threshold (50-60 GeV at small zenith angles) \cite{threshold}.

1ES1426+428 was observed with the MAGIC telescope to search for VHE emission during the multiwavelength campaign carried out in May-June 2008.
MAGIC observed the source in 11 nights, at zenith angles between 16$^{\circ}$ and 34$^{\circ}$ in ON-OFF mode: the telescope pointed towards the source collecting ON data, while the background (OFF data) was estimated from the observations of regions with similar sky conditions as the ON data, but from where no $\gamma$-ray signal is expected.
We excluded from the sample data affected by bad atmospheric conditions. The total amount of final data was corresponding to 14.9 hours of ON and 13.4 hours of OFF data.

\begin{figure*}[th]
  \centering
  \includegraphics[width=5in]{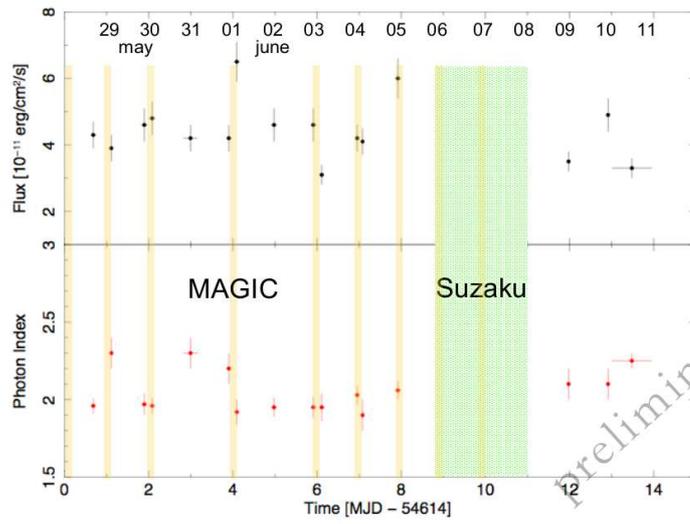}
  \caption{Swift lightcurve for May and June 2008. The MAGIC observation windows are shown together with the Suzaku window. See the text for details.}
  \label{simp_fig}
 \end{figure*}

The collected data were analyzed using the MAGIC standard analysis chain \cite{threshold}, using the Hillas image parameters for parameterizing the images on the camera \cite{Hillas} and timing information of the shower for improving the suppression of light of the night sky \cite{Time}, as well as the hadronic background suppression. It was achieved by means of the Random Forest (RF) method \cite{RF} which allows to compute, for each event, the ``hadronness'' parameter, that is a measure of the probability that the event is not $\gamma$-like. The $\gamma$-ray signal was determined by comparing ON and normalized OFF data in the ALPHA parameter distribution \cite{Hillas}, which, in case of $\gamma$-ray signal, shows an excess at small ALPHA values (below 10$^{\circ}$).

 The analysis did not reveal a significant VHE signal for 1ES1426+428 neither using the whole data set nor considering night by night observations.
Also the search for a signal in different energy bins did not yield a
significant excess. We calculated upper limits in terms of the number of excess events with 95 \% CL using the method described in \cite{UL} and taking into account a systematic error of 30 \% for energy estimation and for effective area calculation \cite{RF}.\\
The number of the excess events was converted into flux upper limits assuming a photon index of -3, which represents an average slope values for this source compared to historical detections (-3.55 in \cite{Whipple}, -3.5 in \cite{VER}, -2.6 in \cite{HEGRA}).

The derived differential upper limits in 4 energy bins are summarized in Table \ref{tableI} (they are not corrected for the IR EBL absorption).
The integral value for the total energy range upper limit is 3 \% of the Crab Nebula flux above 130 GeV.

  \begin{table}[!hb]
 \caption{Flux Upper Limits in May-June 2008}
  \label{tableI}
  \centering
  \begin{tabular}{|c|c|c|}
  \hline
   Threshold energy E$_{th}$(GeV)  & Flux ULs (cm$^{-2}$s$^{-1}$)  & C.U.(\%) \\
   \hline 
    130 & 9.07 10$^{-12}$ & 4 \\
    260 & 2.26 10$^{-12}$ &  2 \\
    518 & 5.57 10$^{-13}$ & 2 \\
    1000 & 2.04 10$^{-13}$ & 1 \\
  \hline 
  \end{tabular}
 The columns represent from left to the right: the threshold energy (which corresponds to peak energies of $\gamma$-ray MonteCarlo samples after all analysis cuts), the flux in units of cm$^{-2}$ s$^{-1}$ and in Crab units (C.U.), based on measurements of the Crab performed with the MAGIC telescope \cite{Crab}.
\end{table}

\subsection{X-ray satellite observations}
 Swift observed the source with short snapshots of 1-2 ks each from May 28 to June 10, overlapping with MAGIC observations when possible.
All Swift observations were performed using all three on-board instruments: the X-ray telescope (XRT), the UV and optical telescope (UVOT) and the Burst Alert Telescope (BAT). In the following only XRT data from 0.2 to 10 keV will be used.

The joint Japanese-US satellite Suzaku \cite{Suzaku} covered the soft and hard X-rays region (0.4-70 keV). The observations were carried out between June 5 to June 8 for a total time of 100 ks. Suzaku data analysis was performed in a similar way as described in \cite{Tagliaf}. A detailed discussion on the used data and analysis will be presented in subsequent publication.

\begin{figure*}[!ht]
  \centering
  \includegraphics[width=5in]{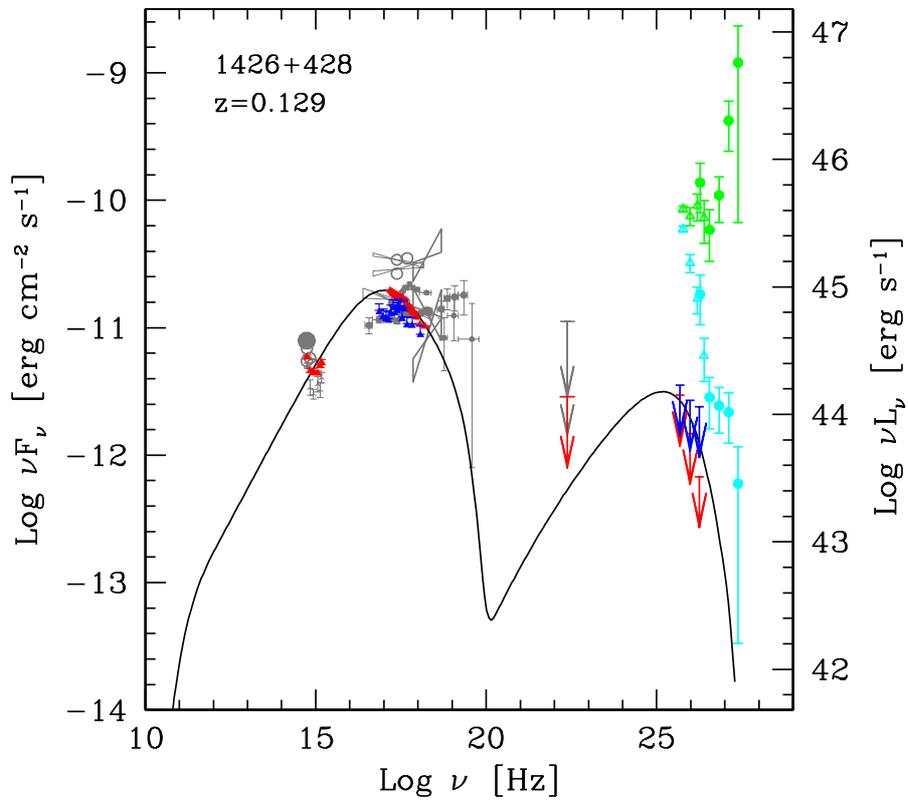}
  \caption{The SED of 1ES 1426+428 as measured during the May-June 2008 MWL campaign together with historical data [PRELIMINARY]. See text for details.}
  \label{sed}
 \end{figure*}

1ES1426+428 was detected both by Suzaku and Swift. Swift data were reduced using the xrtpipeline software distributed with the heasoft 6.3.2 package  by the NASA High Energy Astrophysics Archive Research Center (HEASARC). The X-ray light curve was extracted through the lcurve package. 

The results are reported in Figure \ref{simp_fig} where the Swift light curve is shown in the upper panel together with the MAGIC and Suzaku observation windows. In the lower panel the values of the photon index during the MWL campaign are reported.
The corresponding X-ray flux in the 2-10 keV band is on average 4 $\times$ 10$^{-11}$ erg cm$^{-2}$ s$^{-1}$. Compared to historical data this corresponds to an intermediate activity of the source.

\begin{figure*}[th]
  \centering
  \includegraphics[width=5in]{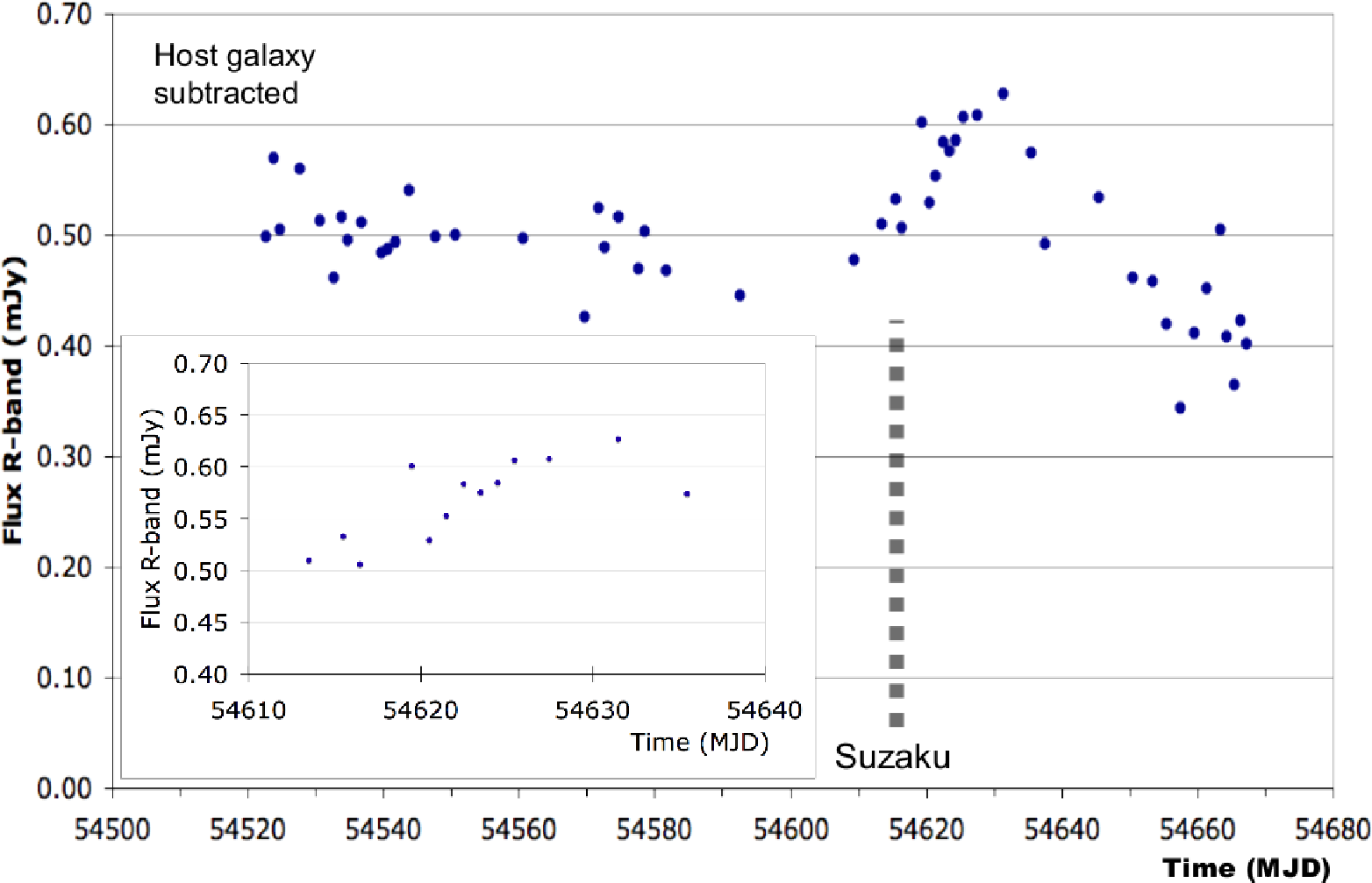}
  \caption{Optical lightcurve of 1ES1426+428 in the R-band, from the KVA telescope. The contribution of the host galaxy is subtracted. See text for details.}
  \label{simp_figII}
 \end{figure*}

\subsection{Optical observations}
The KVA telescope \cite{KVA} is a 35 cm optical telescope located in La Palma. The measurements of the optical flux of 1ES1426+428 were obtained in the Johnson R band. 
In Figure \ref{simp_figII} the optical lightcurve obtained within a 3-months period is shown. The contribution from the host galaxy and nearby stars (F=0.88~mJy) has been subtracted from the overall flux \cite{fluxsub}. The inset shows  the lightcurve in the period of MAGIC observations: the optical flux increased by ~20 \% during the MWL campaign.

\section{Discussion}
The SED of 1ES1426+428 at the epoch of MAGIC observations is shown in Figure \ref{sed} together with historical data. The average Suzaku spectrum is represented in red and the Swift spectra are the red dots for the UV data (taken by UVOT) and the blue dots for the X rays data (taken by XRT). Red and blue arrows report the measured MAGIC upper limits respectively with and without EBL absorption.
 Pale blue and green points are the historical Whipple data \cite{Whipple}, with and without EBL absorption.  The correction for the IR EBL is crucial for determining the correct intrinsic shape of the spectrum. We report for this SED the following parameters: injected electron Lorentz factors: $\gamma$$_{min}$= 10$^2$, $\gamma$$_{max}$= 10$^6$, tangled magnetic field intensity B= 0.18 G, electron number density N= 2.8 $\times$ 10$^3$ particles cm$^{-3}$, emission region dimension R= 5.9 $\times$ 10$^{15}$ cm, Doppler beaming factor D= 25.

 In this campaign the source has shown an intermediate state in the X-ray integral flux whereas the photon index is $>$ 2, thus indicating a synchrotron peak well below 1 keV. At the same time MAGIC could not detect the source on the level of a few percent of Crab. From the comparison with historical data it can be seen that MAGIC observed the source during a low VHE state, while the corresponding X-ray flux is not so low: the absolute value shows a moderate activity, but the slope is quite different from the one obtained with previous observations. The historical X-ray data taken during the VHE outburst show a hardening of the spectrum with photon index $<$ 2 with the synchrotron peak at higher energies, up to 100 keV.

 This suggests that the VHE emission may be related to the slope of the injected electron distribution. Since the X-ray spectrum is in average flat (slope $\simeq$ 2) a minimal change to the electron distribution could produce a dramatic change in the synchrotron peak location.
Further multiwavelength observations could reveal the physical quantities (e.g. the slope or energy break of the injected electrons distribution) that govern this interplay among X-ray and VHE spectral properties.\\

MAGIC has observed 1ES1426+428 during a low state in the VHE band. Upper limits in the 100-1000 GeV range have been derived. The corresponding X-ray spectrum is hard with flux F= 4 $\times$ 10$^{-11}$erg cm$^{-2}$ s$^{-1}$ and the synchrotron peak is located at energies lower than 1 keV. 

The comparison of VHE and X-ray state with historical data, as can be seen in figure \ref{sed}, points to a correlation of the VHE activity with the hardening of the X-ray spectrum and the corresponding shift of the synchrotron peak to higher energies.

\section{Acknowledgments}

We would like to thank the Instituto de Astrofisica de Canarias for the excellent working conditions at the Observatorio del Roque de los Muchachos in La Palma. The support of the German BMBF and MPG, the Italian INFN and Spanish MICINN is gratefully acknowledged. This work was also supported by ETH Research Grant TH 34/043, by the Polish MNiSzW Grant N N203 390834, and by the YIP of the Helmholtz Gemeinschaft. We also thank the KVA observatory and the Suzaku and the Swift teams for providing the data.

 \end{document}